\documentclass[conference]{IEEEtran}
\IEEEoverridecommandlockouts
\usepackage{cite}

\usepackage{amsmath,amssymb,amsfonts}
\usepackage{algorithmic}
\usepackage{graphicx}
\usepackage{textcomp}
\usepackage{xcolor}
\usepackage[acronym]{glossaries}
\usepackage{booktabs}
\usepackage[T1]{fontenc}
\usepackage{times}  

\usepackage[letterpaper, 
            left=0.625in, 
            right=0.625in, 
            top=0.75in, 
            bottom=1.05in, 
            footskip=0in]   
            {geometry}
\usepackage{fancyhdr}
\begin{document}

\title{UAV-Assisted Resilience in 6G and Beyond Network Energy Saving: A Multi-Agent DRL Approach\\}

\author{
	\IEEEauthorblockN{Dao Lan Vy Dinh$^1$, Anh Nguyen Thi Mai$^1$, Hung Tran$^1$, Giang Quynh Le Vu$^{1,6}$, Tu Dac Ho$^2$, \\ 
    Zhenni Pan$^3$, Vo Nhan Van$^4$, 
    Symeon Chatzinotas$^5$, Dinh-Hieu Tran$^5$}  \\
    \IEEEauthorblockA{
    $^1$DATCOM Lab, Faculty of DS\&AI, College of Technology, National Economics University, Vietnam.\\
    $^2$Department of Information Security and Communication Technology, Norwegian University of Science and Technology,\\ Norway.\\
     $^3$Waseda University, Japan.\\
    $^4$Faculty of Information Technology, Duy Tan University, Da Nang 550000, Vietnam.\\    
    $^5$Interdisciplinary Centre for Security, Reliability and Trust (SnT), University of Luxembourg.\\    
    $^{1,6}$Faculty of Information Technology, National Academy of Education Management, Hanoi, Vietnam.\\
    Emails: hung.tran@neu.edu.vn, tu.d.ho@ntnu.no, zhenni.pan@aoni.waseda.jp, 
    \\
     vonhanvan@dtu.edu.vn, symeon.chatzinotas@uni.lu, hieu.tran-dinh@uni.lu, \\ anhntm.datcom@st.neu.edu.vn, vydinhdaolan.datcom@st.neu.edu.vn
    }
    }

\maketitle
\vspace{1mm}

\begin{abstract}
This paper investigates the unmanned aerial vehicle (UAV)-assisted resilience perspective in the 6G network energy saving (NES) scenario. More specifically, we consider multiple ground base stations (GBSs) and each GBS has three different sectors/cells in the terrestrial networks, and multiple cells may become inactive due to unexpected events such as power outages, disasters, hardware failures, or erroneous energy-saving decisions made by external network management systems. During the time required to reactivate these cells, UAVs are deployed to temporarily restore user service. To address this, we propose a Multi-Agent Deep Deterministic Policy Gradient (MADDPG) framework to enable UAV-assisted communication by jointly optimizing UAV trajectories, transmission power, and user-UAV association under a sleeping ground base station (GBS) strategy. This framework aims to ensure the resilience of active users in the network and the long-term operability of UAVs. Specifically, it maximizes service coverage for users during power outages or NES zones, while minimizing the energy consumption of UAVs. Simulation results demonstrate that the proposed MADDPG policy consistently achieves high coverage ratio across different testing episodes, outperforming other baselines. Moreover, the MADDPG framework attains the lowest total energy consumption, while maintaining a comparable user service rate. These results confirm the effectiveness of the proposed approach in achieving a superior trade-off between energy efficiency and service performance, supporting the development of sustainable and resilient UAV-assisted cellular networks.
\end{abstract}

\begin{IEEEkeywords}
6G, centralized training and decentralized execution (CTDE), Multi-Agent Deep Reinforcement Learning,  Network Energy Saving, Self-Organizing Networks (SONs), Unmanned Aerial Vehicle (UAV).
\end{IEEEkeywords}

\section{Introduction}
The rapid proliferation of mobile devices and the shift toward 6G and beyond networks have driven an unprecedented surge in energy consumption. This escalating energy demand not only increases operational expenditure for mobile network operators but also amplifies the environmental impact through higher carbon emissions. Consequently, improving the energy efficiency (EE) and sustainability of wireless infrastructures has become a central design goal for next-generation networks. Beyond energy concerns, modern communication systems must remain resilient to failures caused by natural disasters, backhaul congestion, or equipment malfunctions, necessitating rapid and reliable service recovery under dynamic conditions \cite{b2}.

Unmanned Aerial Vehicles (UAVs) have recently emerged as key enablers for green and resilient communications due to their high mobility, adaptive deployment, and strong line-of-sight (LoS) connectivity. When functioning as Aerial Base Stations (ABSs), UAVs can promptly compensate for service disruptions in Cell Outage Compensation (COC) scenarios or temporarily replace low-traffic Ground Base Stations (GBSs) placed in sleep mode to reduce network energy consumption \cite{b3}. However, jointly determining UAV trajectories, transmit power levels, and the on/off states of GBSs constitutes a complex, non-convex, and temporally coupled optimization problem that challenges traditional model-based methods.

Several works have studied UAV-assisted energy-efficient (EE) architectures via classical optimization. Chang et al. \cite{b11} propose a UAV-GBS scheme for small cells, using a heuristic integer linear program (ILP) to jointly schedule GBS sleep and place UAVs, improving EE while preserving coverage. Alsharoa et al. \cite{b12} optimize a three-tier heterogeneous network (HetNet) with solar-powered drone cells via a per-block binary program, reducing total energy through joint drone deployment and dynamic GBS activation under minimum received-power (QoS) constraints. Li et al. \cite{b3} formulate a unified problem over GBS sleep states, UAV trajectories, and transmit powers; their block-coordinate descent with successive convex approximation (SCA) and branch-and-bound balances flight, transmit, and GBS operational energy. Gaddam et al. \cite{b13} analyze ground radio station infrastructures for UAV links, showing throughput–coverage–EE trade-offs under coordinated scheduling and sleep control.

Collectively, these studies have demonstrated the potential of integrating UAV deployment and GBS sleep mechanisms to achieve greener communication. However, their reliance on deterministic or heuristic optimization techniques, such as ILP, SCA, or block-coordinate methods, limits their scalability and adaptability to highly dynamic network environments. Moreover, none of the above frameworks incorporate learning-based mechanisms that allow agents to autonomously adapt to temporal traffic variations, user mobility, and stochastic wireless conditions.

In contrast, deep reinforcement learning (DRL) has emerged as a powerful approach for complex, non-convex control in wireless networks. Liu et al. \cite{b4} use DRL for energy-efficient UAV coverage control, and Wang et al. \cite{b5} apply Deep Deterministic Policy Gradient (DDPG) to UAV-assisted computation offloading, illustrating DRL's ability to learn adaptive policies. Likewise, \cite{b1} shows that Deep Q-Learning (DQN) can tune GBS configurations for energy savings while preserving QoS. However, few works jointly leverage UAV-assisted architectures and DRL to co-optimize UAV mobility, transmit power, and GBS sleep scheduling within a unified green-communication framework.

To address these limitations, this paper makes four contributions in a single unified framework. First, to the best of our knowledge, it is the first to study UAV-assisted resilience for 6G-and-beyond NES using multi-agent DRL. Second, we model a multi-cell radio access network with scheduled inactive cells, distributed users, and user rate demands with occasional sudden surges. To ensure these surges reflect realistic usage, the traffic patterns are synthesized based on the characteristics of the open 5G dataset from Choi et al. \cite{b15}, which contains real-world traffic from diverse applications such as video conferencing (e.g., Google Meet, MS Teams) and video streaming. Under this model, we aim to maximize user coverage while minimizing UAVs' energy consumption to ensure both network resilience and UAV sustainability. Third, we employ a Multi-Agent Deep Deterministic Policy Gradient (MADDPG) approach with centralized training and decentralized execution to jointly optimize UAV trajectories, transmit power, and user–UAV association. Finally, extensive simulations show that, compared with benchmark schemes, our learning-driven method offers superior adaptation, scalability across heterogeneous environments, and robustness to unpredictable network dynamics.

\section{System Model}\label{sec:formul}
We consider a downlink communication network in which UAVs provide service to ground users when nearby cells are temporarily turned off to save energy or due to power outages or failures. The goal is to simultaneously optimize the UAV trajectory and transmit power to maximize user coverage while minimizing the UAV's energy consumption.

\begin{figure}[!t]    
    \centering
    \includegraphics[width=0.8\linewidth]{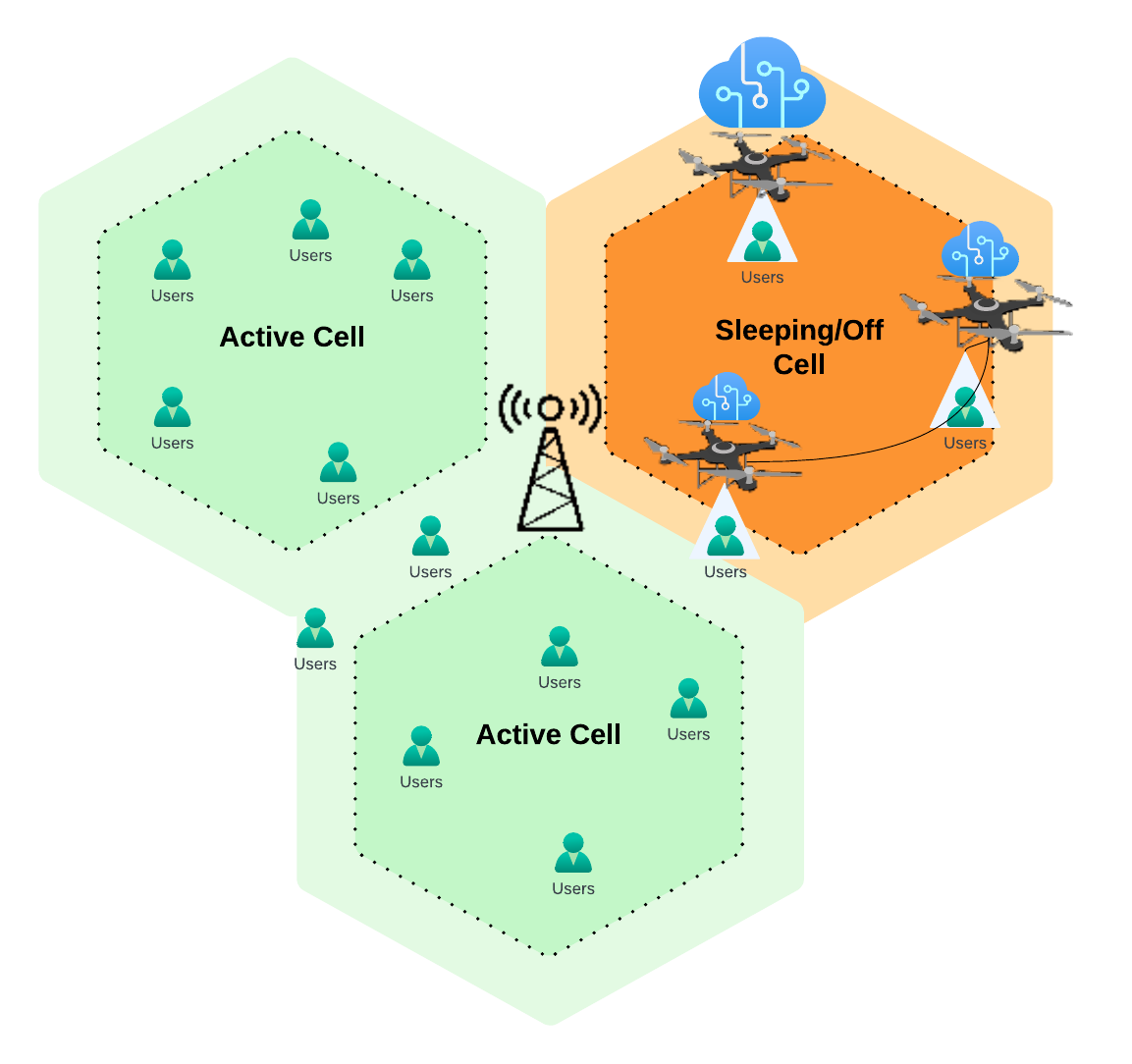}
    \caption{System Model Illustration}
    \label{fig:sysmod}
\end{figure}

As shown in Fig. \ref{fig:sysmod}, each GBS consists of three sectors or cells. Each cell serves the associated mobile users within its coverage area. Cells may become inactive for several reasons, including unexpected events such as power outages, hardware failures, or natural disasters. In addition, modern networks often employ automated energy-saving mechanisms that switch base stations into sleep mode based on traffic predictions. However, these machine-learning-based decisions may occasionally be inaccurate, causing cells to remain in sleep mode even when user demand suddenly increases. Since reactivating a sleeping base station may take several minutes, users in these areas may experience temporary service disruption. Even when turned on, these cells are still overloaded and cannot serve all users in time. To address this gap, UAVs acting as aerial base stations are deployed to provide immediate temporary coverage. We design the scheduling so that users are allocated to the UAV with the best signal-to-interference-plus-noise ratio (SINR). This ensures that each user is connected to the most favorable serving UAV based on the instantaneous channel quality and the UAV's location.

\subsubsection{Cells}
There are $K$ fixed cells, each is a hexagonal cell. The operational state of each cell $k$, denoted by $s_k(t) \in \{0,1\}$, is determined by external network conditions or network management policies, such as energy-saving mechanisms, unexpected outages, or emergency situations. These states are treated as environment inputs rather than optimization variables in this work. Specifically, to minimize the substantial static power consumption of GBSs, a cell is switched to sleep mode ($s_k(t)=0$) if the aggregated traffic demand of its associated users falls below a predefined energy-efficiency threshold $\gamma_{th}$. Conversely, the cell remains active ($s_k(t)=1$) during peak loads. This pre-optimized sleeping schedule serves as the baseline environment input for the subsequent UAV resilience optimization.

\subsubsection{UAVs}
There are $N$ UAVs in the network with each UAV $i$ flies at a fixed altitude H with position $q_i(t) = 
    [x_i(t),y_i(t)]$. UAVs can move within a bounded area, with $\Delta_t =1$, movement constrained by: 
    \begin{equation} \label{eq:v_max}
        ||\Delta q_i(t)|| \leq v_{max}.
    \end{equation} 
    Each UAV has limited transmit power:
    \begin{equation} \label{eq:pi}
        0 \leq P_i(t) \leq P_{max}.
    \end{equation} The total system transmit power throughout flying time must satisfy: \begin{equation} \label{eq:p_max}
        \sum_{i} P_i(t) \leq P_{max}.
    \end{equation}

\subsubsection{Users}
$M$ users are distributed randomly within the coverage area. Each user $j$ has a fixed location $(x_j,y_j)$ and a heterogeneous, dynamic rate requirement $R_{req,j}$.

\subsection{Communication Model}
The formulation for the communication model, including the distance,
Rician fading channel, SINR, and achievable data rate 
(Eqs.~\eqref{distance} -- \eqref{achieve_rate}), 
is adopted from the model presented in \cite{b10,TranUAV2022}.

Given the location $(x_j,\ y_j,\ 0)$ of each user $j \in \mathcal{M}$ at ground level and the current UAV location $(x_i,\ y_i,\ H)$ at timestep $t$, the 3D Euclidean distance $d_{ij}(t)$ between UAV $i$ and user $j$ is:
\begin{equation}
    d_{ij}(t) = \sqrt{H^2+(x_i(t)-x_j)^2+(y_i(t)-y_j)^2}.
    \label{distance}
\end{equation}

The channel coefficient between user $j$ and UAV $i$ follows a Rician fading model:
\begin{equation}
    h_{ij}(t)=\sqrt{w_{ij}(t)} \tilde{h}_{ij}(t),
    \label{channel_coef}
\end{equation}
where $w_{ij}(t)$ represents large-scale path loss:
\begin{equation}
    w_{ij}(t)=w_0 d_{ij}^{-\alpha}(t),
    \label{large_scale}
\end{equation}
with $w_0$ being the reference path loss at $d=1$m and $\alpha\ge 2$ the path loss exponent. The small-scale Rician fading coefficient $\tilde{h}_{ij}(t)$ is given by:
\begin{equation}
\tilde{h}_{ij}(t) =\sqrt{\frac{K}{1+K}}\bar{h}_{ij}(t) + \sqrt{\frac{1}{1+K}}\hat{h}_{ij}(t),
\label{small_scale}
\end{equation}
where $\bar{h}_{ij}(t)$ and $\hat{h}_{ij}(t)\sim \mathcal{CN}(0,\,1)$ denote the deterministic line-of-sight (LoS) and non-line-of-sight (NLoS) Rayleigh fading components, respectively, and $K$ is the Rician K-factor.

SINR experienced by user $j$ when served by UAV $i$ is:

\begin{equation}
\gamma_{ij}(t) =
\frac{P_i(t) |h_{ij}(t)|^2}{
\sigma^2 + \sum_{k \neq i}^{N} P_k(t) |h_{kj}(t)|^2
+ \sum_{\ell=1}^{K} s_\ell(t) P_{\text{BS}} |h_{\ell j}(t)|^2
}
\label{sinr}
\end{equation}
where $P_i(t)$ is the transmit power of UAV $i$, $\sigma^2$ is the thermal noise power, the first summation accounts for inter-UAV interference from other active UAVs, and the second summation accounts for interference from active ground base stations (GBSs), with $s_\ell(t) \in \{0,1\}$ indicating whether cell $\ell$ is active and $P_{\text{BS}}$ being the GBS transmit power.

The achievable data rate for user $j$ served by UAV $i$ is:
\begin{align}
    R_{ij}(t) = B\log_2(1+\gamma_{ij}(t)),
    \label{achieve_rate}
\end{align}
where $B$ is the allocated bandwidth.

\textbf{User-UAV Association:} Each user is associated with the UAV providing the highest SINR among all active UAVs:
\begin{equation}
a_{ij}(t) =
\begin{cases}
1, & \text{if } i = \displaystyle \arg\max_{k \in \mathcal{N}} \gamma_{kj}(t) \text{ and } s_{\text{cell},j}(t) = 0, \\
0, & \text{otherwise,}
\end{cases}
\end{equation}
where $s_{\text{cell},j}(t) = 0$ indicates that user $j$'s home cell is in sleep mode (requiring UAV service). Each user can be served by at most one UAV:
\begin{equation}\label{eq:served_user}
    \sum_{i=1}^{N} a_{ij}(t) \le 1, \quad \forall j \in \mathcal{M}.
\end{equation}

\textbf{QoS Constraint:} To ensure satisfactory service quality, each served user must achieve a minimum data rate:
\begin{equation}
    R_{ij}(t) \ge R_{\mathrm{req},j}(t), \quad \forall j: a_{ij}(t)=1,
    \label{eq:rate_constraint}
\end{equation}
where $R_{\mathrm{req},j}(t)$ is the time-varying rate requirement of user $j$, synthesized based on real-world 5G traffic patterns from \cite{b15}.

\subsection{UAV Energy Model}
The energy consumption of UAV $i$ at time $t$ consists of both propulsion and communication components:
\begin{equation}
    E_i(t) = E_{\text{prop},i}(t) + E_{\text{comm},i}(t),
\end{equation}
where $E_{\text{prop},i}(t)$ and $E_{\text{comm},i}(t)$ are formulated, respectively, as follows:
\begin{align}
    &E_{\text{prop},i}(t) = \alpha_1 ||\Delta q_{i}(t)||^2 + \alpha_2 ||\Delta q_{i}(t)||,
    \\
    &E_{\text{comm},i}(t) = P_i(t) \Delta t,
\end{align}
with $\Delta t$ is a timestep duration.
Accordingly, the total energy of all UAVs is defined as:
\begin{align}
    E_\text{UAV}(t) = \sum_{i=1}^{n} E_i(t).
\end{align}


\subsection{Problem Formulation}


The optimization objective is to maximize user coverage (percentage of served users meeting QoS) while self-awaring total UAV energy consumption over a finite horizon $T$. Formally:
\begin{align}
    \label{eq:P1}
\mathcal{P}_1: &\max_{\{\mathbf{v}_i(t), P_i(t)\}_{i,t}} \quad 
    \sum_{t=1}^{T} C(t) \\
    \text{s.t.}~
    & R_{ij}(t) \ge R_{\mathrm{req},j}, \quad \forall i,t,j, \IEEEyessubnumber\label{eq:P1:rate}\\
    & \sum_i a_{ij}(t) \le 1, \quad \forall t,j,\label{eq:P1:assoc}\\
    & a_{ij}(t) \in \{0,1\}, \quad \forall i,t, j\label{eq:P1:binary}\\
    & ||\Delta v_i(t)|| \le v_{max}, \quad \forall i,t, \IEEEyessubnumber\label{eq:P1:velocity}\\
    & 0 \le P_i(t) \le P_{max}, \quad \forall i,t, \IEEEyessubnumber\label{eq:P1:power}\\
    & \sum_i P_i(t) \le P^{max}_{\sum}, \quad \forall t, \IEEEyessubnumber\label{eq:P1:sum_power}
\end{align}


where $C(t) = \sum_i C_i(t)$ with $C_i(t)$ is user coverage per UAV, defined in reward function in section \ref{sec:pomdp}. Constraint~\eqref{eq:P1:rate} ensures that each served user meets its minimum QoS requirement, while~\eqref{eq:P1:assoc} guarantees that each user is associated with at most one UAV. Constraints~\eqref{eq:P1:velocity}-\eqref{eq:P1:sum_power} enforce UAV mobility and power limitations.


\section{POMDP Formulation for UAV-Assisted Resilience in 6G NES}\label{sec:pomdp}

We model the multi-UAV control problem as a partially observable Markov decision process (POMDP) \cite{b6} with CTDE. During training, a global state $s_{\text{global}}(t)$ is available to all agents' critics. During execution, each UAV $i$ operates based solely on its local observation $o_i(t)$.

\subsection{State and Observation Spaces}

\subsubsection{Local Observation of UAV $i$}
Each UAV $i$ constructs a local observation vector comprising:
\begin{align}
    o_i(t) = \Big[ & \underbrace{\mathbf{q}_i(t), P_i(t)}_{\text{own state (3D)}}, 
    \underbrace{\{\mathbf{q}_k(t), P_k(t)\}_{k \neq i}}_{\text{partner UAV states}}, \notag \\
    & \underbrace{\{x_j, y_j, R_{\text{req},j}(t)\}_{j \in \mathcal{U}_i(t)}}_{\text{5 nearest users}}, 
    \underbrace{\{s_k(t)\}_{k=1}^{K}}_{\text{cell states}} \Big],
\end{align}
where $\mathcal{U}_i(t) \subset \mathcal{M}$ denotes the set of the 5 nearest users to UAV $i$ (within a radius threshold), limited to reduce observation dimensionality.

\subsubsection{Global State}
For centralized training, the global state aggregates information about all users:
\begin{equation}
    s_{\text{global}}(t) = \Big[ \{x_j, y_j, R_{\text{req},j}(t)\}_{j=1}^{M}, \{s_k(t)\}_{k=1}^{K} \Big]
\end{equation}

\subsection{Action Space}
The continuous action of UAV $i$ at time $t$ is:
\begin{equation}
    \mathbf{a}_i(t) = [v_{x,i}(t), v_{y,i}(t), P_i(t)] \in \mathbb{R}^3,
\end{equation}
where:
\begin{itemize}
    \item $v_{x,i}(t), v_{y,i}(t) \in [-v_{\max}, v_{\max}]$: velocity components (m/s),
    \item $P_i(t) \in [0, P_{\max}]$: transmit power (Watts).
\end{itemize}


\subsection{Reward Function}
The per-timestep reward for UAV $i$ balances coverage contribution and energy penalty:
\begin{equation}
    r_i(t) = \omega_1 \cdot \mathcal{C}_i(t) - \omega_2 \cdot \frac{E_i(t)}{E_{\max}},
    \label{reward_func}
\end{equation}
where $\omega_1 >> \omega_2$, and $\mathcal{C}_i(t)$, $E_i(t)$ and $E_{max}$ are calculated, respectively, as:
\begin{itemize}
    \item $\mathcal{C}_i(t) = \frac{|\{j : a_{ij}(t)=1, R_{ij}(t) \ge R_{\text{req},j}(t)\}|}{|\{j : s_{\text{cell},j}(t)=0\}|}$,
    \item $E_i(t) = P_i(t) \Delta t + \alpha_1 \|\mathbf{v}_i(t)\|^2 + \alpha_2 \|\mathbf{v}_i(t)\|$,
    \item $E_{\max} = P_{max}\Delta t + \alpha_1 \|\mathbf{v}_{max}(t)\|^2 + \alpha_2 \|\mathbf{v}_{max}(t)\|$
\end{itemize}

\subsection{Learning Objective}
Each UAV learns a stochastic policy $\pi_i(\mathbf{a}_i | o_i)$ to maximize expected cumulative discounted return:
\begin{equation}
    J(\pi_1, \ldots, \pi_N) = \mathbb{E}_{\tau \sim \pi} \left[ \sum_{t=0}^{T} \gamma^t r_i(t) \right],
\end{equation}
where $\gamma$ is the discount factor and $\tau$ denotes a trajectory.

\section{The Proposed CTDE-MADDPG Framework}
\label{sec:method}

\subsection{CTDE-MADDPG Framework}
\begin{figure}
    \centering
    \includegraphics[width=0.8\linewidth]{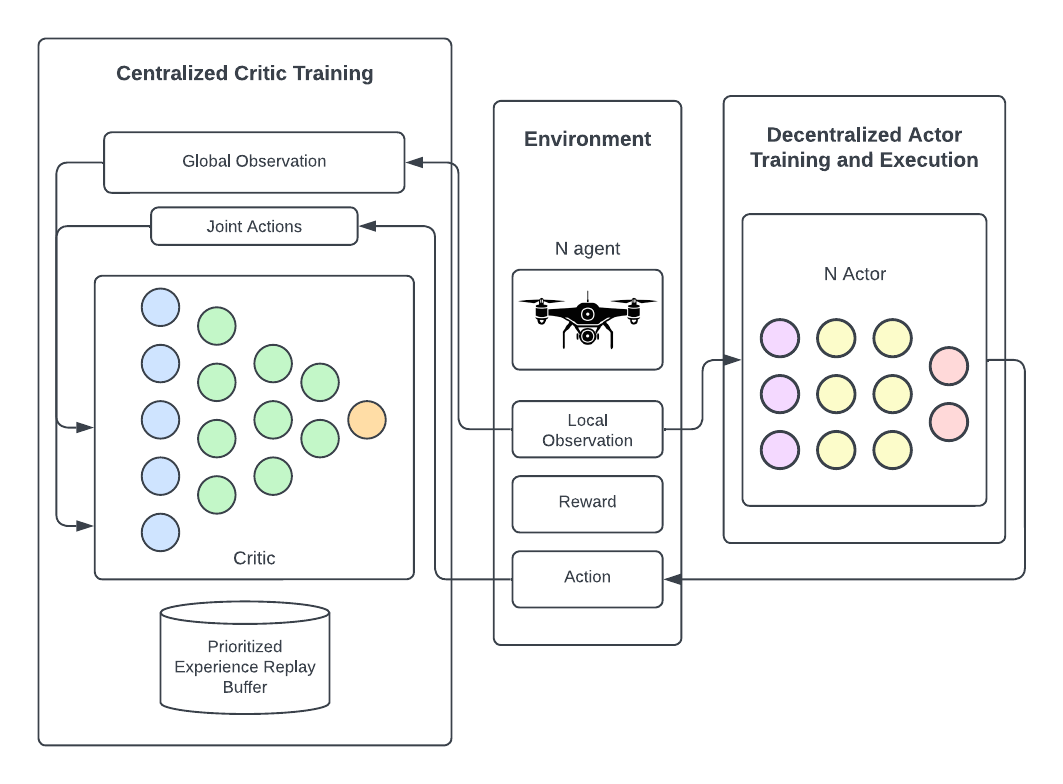}
    \caption{CTDE-MADDPG Framework}
    \label{fig:ctde}
\end{figure}

We adopt a CTDE variant of MADDPG for cooperative UAVs control with better memory and training efficiency as shown \ref{fig:ctde}. Each UAV is modeled as an agent with its own actor–critic pair. During training, each agent's critic has access to the global state $s_\text{global}(t)$ and actions of all agents, enabling coordinated learning. However, during execution, each UAV only uses its local observations $s_i(t)$ through the actor network, making the system scalable and robust to partial observability.

\subsection{Network Architecture}
Both actor and critic networks are implemented as fully connected multilayer perceptrons (MLPs) with ReLU activations. Actor outputs continuous actions $[v_x, v_y, P]$ normalized to valid ranges via tanh activation, with $v_x,v_y \in [-v_{max}, v_{max}]$ and $P \in [0,P_{max}]$. Critic receives concatenated joint observations and actions. Optimization is performed with the Adam optimizer and gradient norms are clipped to prevent instability. Target networks are softly updated using Polyak averaging.

\subsection{Training Mechanism}
\subsubsection{Prioritized Experience Replay (PER)}
Standard experience replay samples transitions uniformly, which can be inefficient when most experiences are not informative. To improve sample efficiency and stability, we integrate a prioritized replay buffer based on the proportional prioritization scheme. Each transition’s sampling probability is determined by its temporal-difference (TD) error magnitude, stored in a sum-tree structure for efficient $O(logN)$ sampling and priority updates.

\subsubsection{Exploration Strategy}
Besides dynamically adjusting exploration through PER, we also add temporally correlated Ornstein-Uhlenbeck (OU) Noise to actions during training to encourage stable exploration:
$N_t = N_{t-1} + \theta(\mu - N{t-1}) \Delta t + \sigma W_t$
with noise parameter $\sigma$ is decayed by 0.9999 per step with a minimum of 0.01 \cite{b7}. After substantial number of episodes, exploration noise is disabled to exploit the learned policy. This prevents erratic movements that could violate physical constraints, and produces smooth, correlated actions suitable for UAV dynamics.

\subsubsection{Interference Management}
The system accounts for inter-UAVs interference when computing SINR values. Each UAV’s transmission contributes to the interference perceived by users connected to other UAVs. This coupling is reflected in the environment dynamics and achievable rate calculation. The mechanism encourages UAV agents to learn cooperative spatial and power control strategies that balance coverage maximization and energy minimization.

\section{Results and Discussions}\label{sec:result}

\subsection{Reward-based Analysis}
\begin{figure}[ht]
    \centering
    \includegraphics[width=0.8\linewidth]{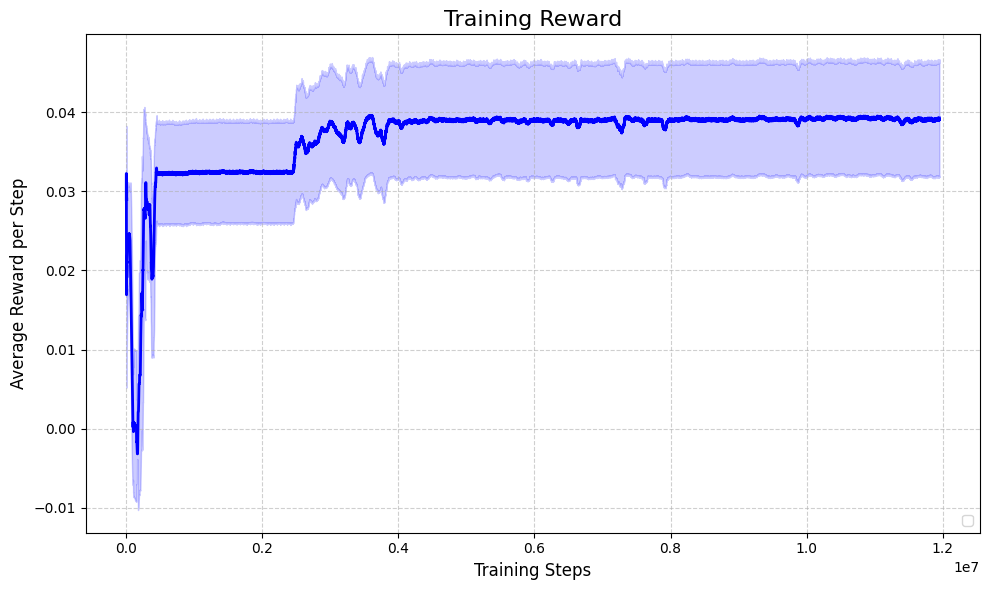}
    \caption{Training convergence curve, showing the average reward per step over the training process.}
    \label{fig:train_reward}
\end{figure}

Fig.~\ref{fig:train_reward} illustrates the convergence of the proposed MADDPG framework. The x-axis represents the Training Episodes (up to 25,000), and the y-axis is the normalized average episodic reward (per step). The average reward rises sharply during the initial exploration phase and stabilizes around 0.032. It exhibits a clear stepped improvement around the 5,000-episode mark, which coincides with the learning-rate decay. This jump is followed by a brief period of high-variance fine-tuning. The policy then settles into its final converged state, and the stable plateau from roughly 8,000 episodes onward confirms robust convergence.

\begin{figure}[ht]
    \centering
    \includegraphics[width=0.8\linewidth]{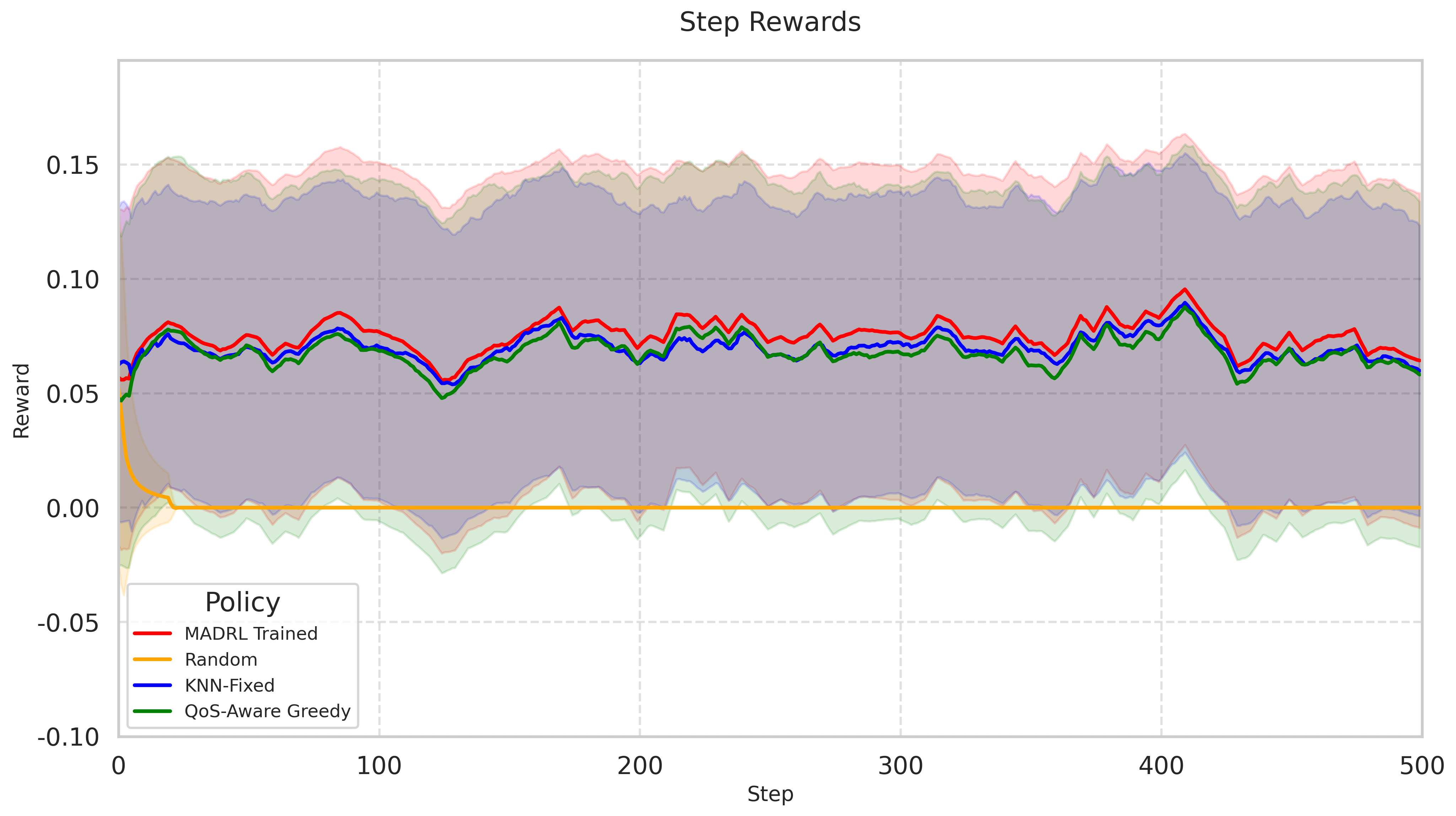}
    \caption{Comparison of instantaneous reward per step among four policies, averaged over 15 testing episodes.}
    \label{fig:test_reward}
\end{figure}

As shown in Fig.~\ref{fig:test_reward}, we compare the instantaneous reward per step of the learned MADDPG policy against three baselines: KNN-Fixed, QoS-Aware Greedy, and Random. The results demonstrate that the MADDPG policy (red line) consistently maintains the highest average reward across all timesteps. 

Specifically, the QoS-Aware Greedy policy (green line) shows competitive performance in the early steps but fluctuates significantly. Although it achieves high instantaneous coverage, its myopic focus on immediate SINR leads to excessive power consumption and inefficient trajectories, thereby reducing the total reward. The KNN-Fixed method (blue line) performs consistently but lacks the adaptability to dynamic traffic changes. In contrast, the MADDPG agents, through cooperative learning, achieve a superior balance between spatial positioning and energy efficiency, enabling them to sustain higher rewards throughout the episode. The Random policy serves as a lower bound, quickly collapsing toward zero reward.

\subsection{Coverage Performance Evaluation}

\begin{figure}[ht]
    \centering
    \includegraphics[width=0.8\linewidth]{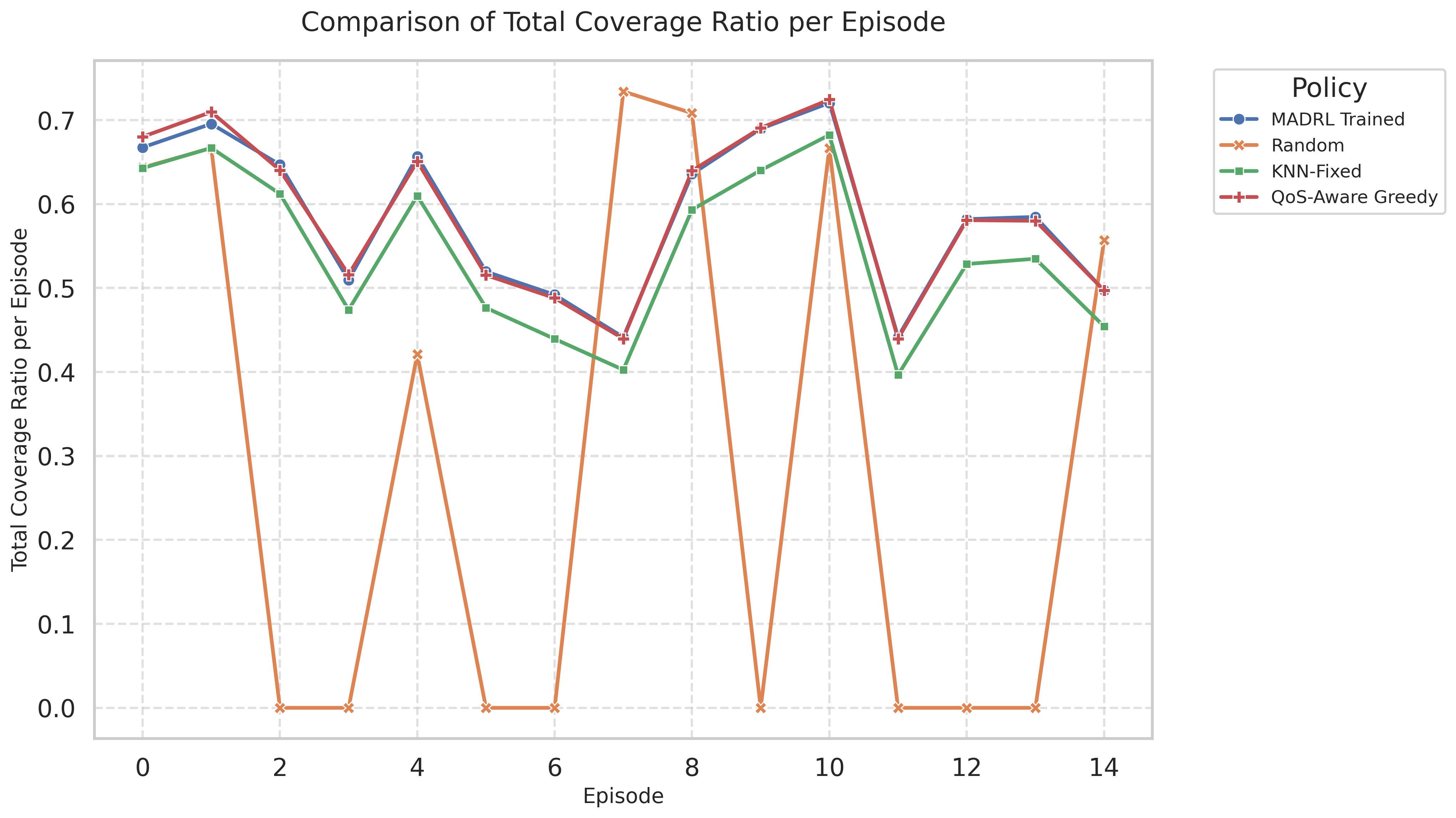}
    \caption{Comparison of Total Coverage Ratio per Episode.}
    \label{fig:cov_plot}
\end{figure}

Fig. \ref{fig:cov_plot} illustrates the total coverage ratio per episode. The QoS-Aware Greedy policy achieves the highest instantaneous coverage ($78.42\%$) in most episodes by aggressively directing UAVs toward signal peaks. However, our proposed MADDPG framework achieves a highly comparable coverage rate ($78.37\%$), with a negligible difference of approx. $0.05\%$. This indicates that MADDPG successfully learns to identify and cover critical user clusters almost as effectively as an exhaustive greedy search. 

Both algorithms significantly outperform KNN-Fixed ($76.20\%$) and Random, confirming the necessity of intelligent trajectory planning in dynamic sleeping cell scenarios. The performance gap between MADDPG/Greedy and KNN highlights that simply moving to the nearest user (KNN) often leads to suboptimal positioning that neglects cluster-edge users, whereas optimization-based approaches provide better spatial distribution.

\section{Conclusion}
This paper presents a UAV-assisted resilience framework for 6G energy-saving networks using a CTDE-based MADDPG approach. The proposed method enables UAVs to autonomously adapt their trajectories, transmit power, and user association to dynamically compensate for coverage holes induced by traffic-aware GBS sleeping strategies. Simulation results demonstrate that our framework achieves a superior trade-off between resilience and sustainability, attaining user coverage rates comparable to an aggressive QoS-Aware Greedy baseline while consuming drastically less energy, saving approximately 24\% compared to the conventional All-Cell-ON configuration. These findings highlight that while heuristic greedy strategies can effectively restore coverage, they are energetically unsustainable for UAV-assisted networks due to excessive propulsion costs. Future work includes lightweight policy distillation to speed convergence, integrating dynamic bandwidth allocation to better satisfy per-user QoS, and jointly optimizing GBS sleep schedules. Overall, this study confirms that integrating MADRL with UAV architectures is a key enabler for autonomous, green, and resilient 6G systems.

\end{document}